\documentclass[sigconf]{acmart}

\usepackage{balance}
\AtBeginDocument{%
  \providecommand\BibTeX{{%
    \normalfont B\kern-0.5em{\scshape i\kern-0.25em b}\kern-0.8em\TeX}}}
    
\copyrightyear{2020}
\acmYear{2020}
\setcopyright{acmlicensed}
\acmConference[HRI '20]{Proceedings of the 2020 ACM/IEEE International Conference on Human-Robot Interaction}{March 23--26, 2020}{Cambridge, United Kingdom}
\acmBooktitle{Proceedings of the 2020 ACM/IEEE International Conference on Human-Robot Interaction (HRI '20), March 23--26, 2020, Cambridge, United Kingdom}
\acmPrice{15.00}
\acmDOI{10.1145/3319502.3378178}
\acmISBN{978-1-4503-6746-2/20/03}

\usepackage{algorithm}
\usepackage{algorithmic}
\usepackage{subcaption}
\usepackage{amssymb}
\usepackage{xcolor}
\usepackage{soul}

\usepackage[framemethod=TikZ]{mdframed}
\mdfdefinestyle{MyFrame}{%
    linecolor=black,
    outerlinewidth=0.5pt,
    innertopmargin=4pt,
    innerbottommargin=4pt,
    innerrightmargin=4pt,
    innerleftmargin=4pt,
        leftmargin = 4pt,
        rightmargin = 4pt
        }

\begin{document}
\fancyhead{}

\title{Four Years in Review: Statistical Practices of Likert Scales in Human-Robot Interaction Studies}

\author{Mariah L. Schrum}
\authornotemark[1]
 \email{mschrum3@gatech.edu}
 \affiliation{
   \institution{Georgia Institute of Technology}
   \streetaddress{266 Ferst Dr NW}
   \city{Atlanta}
   \state{Georgia}
   \postcode{30332}
}

\author{Michael Johnson}
\authornotemark[1]
\email{michael.johnson@gatech.edu}
 \affiliation{
   \institution{Georgia Institute of Technology}
   \streetaddress{266 Ferst Dr NW}
   \city{Atlanta}
   \state{Georgia}
   \postcode{30332}
}

\author{Muyleng Ghuy}
\authornote{All three authors contributed equally to this research.}
\email{mghuy3@gatech.edu}
 \affiliation{
   \institution{Georgia Institute of Technology}
   \streetaddress{266 Ferst Dr NW}
   \city{Atlanta}
   \state{Georgia}
   \postcode{30332}
}

\author{Matthew C. Gombolay}
\email{matthew.gombolay@cc.gatech.edu}
 \affiliation{
   \institution{Georgia Institute of Technology}
   \streetaddress{266 Ferst Dr NW}
   \city{Atlanta}
   \state{Georgia}
   \postcode{30332}
}


\begin{abstract}
  As robots become more prevalent, the importance of the field of human-robot interaction (HRI) grows accordingly. As such, we should endeavor to employ the best statistical practices. Likert scales are commonly used metrics in HRI to measure perceptions and attitudes. Due to misinformation or honest mistakes, most HRI researchers do not adopt best practices when analyzing Likert data. We conduct a review of psychometric literature to determine the current standard for Likert scale design and analysis. Next, we conduct a survey of four years of the International Conference on Human-Robot Interaction (2016 through 2019) and report on incorrect statistical practices and design of Likert scales. During these years, only 3 of the 110 papers applied proper statistical testing to correctly-designed Likert scales. Our analysis suggests there are areas for meaningful improvement in the design and testing of Likert scales. Lastly, we provide recommendations to improve the accuracy of conclusions drawn from Likert data. 
\end{abstract}

\ccsdesc[500]{General and reference~Surveys and overviews}
\ccsdesc[300]{General and reference~Evaluation}
\ccsdesc[100]{General and reference~Metrics}


\keywords{Metrics for HRI; Likert Scales; Statistical Practices}

\maketitle

\section{Introduction}
The study of human-robot interaction is the interdisciplinary examination of the relationship between humans and robots through the lenses of psychology, sociology, anthropology, engineering and computer science. This all-important intersection of fields allows us to better understand the benefits and limitations of incorporating robots into a human's environment. As robots become more prevalent in our daily lives, HRI research will become more impactful on robot design and the integration of robots into our societies. Therefore, it is critical that best scientific practices are employed when conducting HRI research.

\begin{figure}[b]
\centering
    \includegraphics[width=0.8\linewidth]{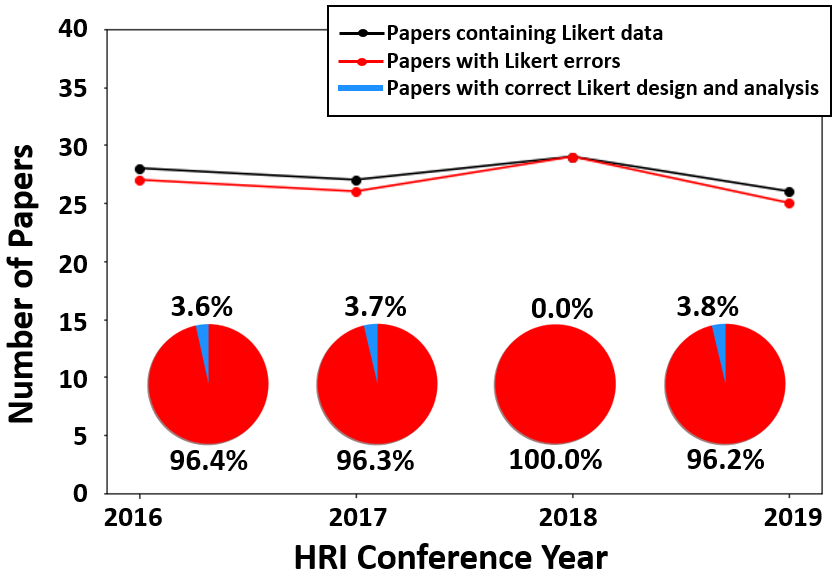}
    \caption{An overview of HRI proceedings with different types of errors when handling Likert data from 2016 - 2019. }
    \label{fig:error_overview}
\end{figure}

Likert scales, a commonly employed technique in psychology and more recently in HRI, are used to determine a person's attitudes or opinions on a topic~\cite{Likert1932}. Statistical tests can then be applied to the responses to determine how an attitude changes between different treatments. Such studies provide important information for how best to design robots for optimal interaction with humans. Because of the nearly universal confusion surrounding Likert scales, improper design of Likert scales is not uncommon~\cite{gombolay2016appraisal}. Furthermore, care must be taken when employing statistical techniques to analyze Likert scales and items. Because of the ordinal nature of the data, statistical techniques are often applied incorrectly, potentially resulting in an increased likelihood of false positives. Unfortunately, we find the misuse of Likert questionnaires to occur frequently enough to be worth investigating.

In this paper, we 1) review the psychometric literature of Likert scales, 2) analyze the past four years of HRI papers, and 3) posit recommendations for best practices in HRI. Based upon our review of psychometric literature, we find that only 3 of 110 papers in the last four years of proceedings of HRI research properly designed and tested Likert scales. A summary of our analysis is depicted in Fig.~\ref{fig:error_overview}. Unfortunately, this potential malpractice may suggest that the findings in $97.3$\% of HRI papers that based their conclusions off of Likert scales may warrant a second look.

Our first contribution is comprised of a survey of the latest psychometric literature regarding the current best practices for design and analysis of Likert scales. In cases where there is dissent or disagreement, we present both perspectives. Nonetheless, we find areas of consensus in the literature to establish recommendations for how to best design Likert scales and to analyze their data. In areas of agreement, we provide recommendations to the HRI community for how we can best construct and analyze Likert data.

Our second contribution is a survey of the proceedings of HRI 2016 through 2019 based upon the established best practices. Our review revealed that a majority of papers incorrectly design Likert scales or improperly analyze Likert data. Common mistakes are not including enough items, analyzing individual Likert items, not verifying the assumptions of the statistical test being applied, and not performing appropriate post-hoc corrections. 

Our third and final contribution is a discussion of how we, as a field, can correct these practices and hold ourselves to a higher standard. Our purpose is not to dictate legalistic rules to be followed at penalty of a paper rejection. Instead, we seek to open up the floor for a constructive debate regarding how we can best establish and abide by our agreed upon best practices in our field. We hope that in doing so, HRI will continue to have a strong, positive influence on how we understand, design, and evaluate robotic systems.

\vspace{4pt}
\begin{mdframed}[style=MyFrame]
\textbf{Nota Bene:} \textit{We confess we have not employed best practices in our own prior work. Our goal for this paper is not to disparage the field, but instead to call out the ubiquitous misuse of a vital metric: Likert scales. We hope to improve the rigor of our own and others' statistical testing and questionnaire design so that we can stand more confidently in the inferences  drawn from these data.}
\end{mdframed}


\section{Literature Review \& Best Practices}
Likert scales play a key role in the study of human-robot interaction. Between 2016 and 2019, Likert-type questionnaires appeared in more than 50\% of all HRI papers.  As such, it is imperative that we make proper use of Likert scales and are careful in our design and analysis so as not to de-legitimize our findings. We begin with a literature review to investigate the current best practices for Likert scale design and statistical testing. We acknowledge that reviews concerning the design and analysis of Likert scales have been previously conducted \cite{Subedi2016,Carifio2007,Jamieson2004}. However, our analysis is the first targeted at the HRI community, and we believe it is important to ground our discussion in the current understanding of the best methods related to the construction and testing of Likert data as found in the psychometric literature. 

Many of the debates surrounding Likert scale design and analysis are unsettled.  As such, we present both sides of these arguments and reason through the areas of agreement and disagreement to arrive at our own recommendations for how HRI researchers can best navigate these often murky waters.

\subsection{What is a Likert Scale?} \label{LikertScale}
\begin{figure}[h!]
    \centering
    \includegraphics[width=\linewidth]{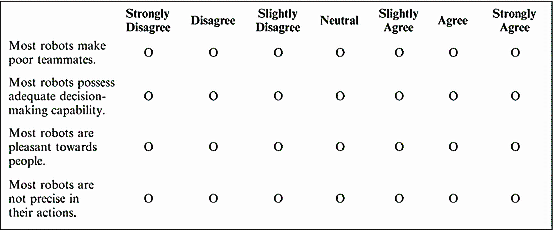}
    \caption{This figure illustrates a portion of a balanced Likert scale measuring trust (Courtesy of \cite{Mittu2016}).}
    \label{fig:LikertScale}
\end{figure}

Likert scales were created in 1932 by Rensis Likert and were originally designed to scientifically measure attitude~\cite{Likert1932}. A Likert scale is defined as "a set of statements
(items) offered for a real or hypothetical situation
under study" in which an individual must choose their level of agreement with a series of statements \cite{Joshi2015}.  The original response scale for a Likert item ranged from one to five (strongly disagree to strongly agree). A seven-point scale is also common practice. An example Likert scale is shown in Fig.~\ref{fig:LikertScale}.

Confusion often arises around the term "scale." A Likert scale does not refer to a single prompt which can be rated on a scale from one to $n$ or "strongly disagree" to "strongly agree". Rather, a Likert scale refers to a set of related prompts or "items" whose individual scores can be summed to achieve a composite score quantifying a participant's attitude toward a latent, specific topic \cite{Carifio2008}. "Response format" is the more appropriate term when describing the options ranging from "strongly disagree" to "strongly agree" \cite{Carifio2007}. This distinction is important for the following reasons. First, a high degree of measurement error arises when a participant is asked to respond only to a single prompt; however, when asked to respond to multiple prompts, this measurement error tends to average out. Second, a single item often addresses only one aspect or dimension of a particular attitude, whereas multiple items can report a more complete picture \cite{Nunnally1994,Woollins1992}. Therefore, it is important to distinguish whether there are multiple items in the scale or simply multiple options in the response format. \cite{Carifio2007} emphasizes the importance of this distinction by stating that the meaning of the term scale "is so central to accurately understanding a Likert scale (and other scales and psychometric principles as well) that it serves as the bedrock and the conceptual, theoretical and empirical baseline from which to address and discuss a number of key misunderstandings, urban legends and research myths.”

It is not uncommon in HRI, as well as psychometric literature, for a researcher to report that he or she employed a five-item Likert scale when in reality he or she used a single item Likert scale with five response options. To ground this distinction in an example, Fig.~\ref{fig:LikertScale} depicts a Likert Scale with four Likert items with seven-option response format. To avoid such confusion, it is important to be precise when describing a Likert scale as a five-option response format has a very different meaning from a five-item Likert scale. Furthermore, a set of items that prompts the user to select a rating on a bipolar scale of antonyms, i.e., human-like to machine-like, is not a true Likert scale. This is a semantic differential scale and should be referred to as such \cite{Verhagen2015}.

\textit{Recommendation - We recommend that HRI researchers be deliberate when describing Likert response formats and scales to avoid confusion and misinterpretation.}

\subsection{Design} 
\label{design}
Because HRI is a relatively new field, HRI researchers often explore novel problems for which they appropriately need to craft problem-specific scales. However, care must be taken to correctly design and assess the validity of these scales before utilizing them for research. The design of the scale is one of the least agreed upon topics pertaining to Likert questionnaires in the psychometric literature.  Disagreement arises around the optimal number or response choices in an item, the ideal number of items that should comprise a scale, whether a scale should be balanced, and whether or not to include a neutral midpoint. Below, we address each topic.

\medskip

\noindent\textbf{Number of Response Options -}
Rensis Likert himself suggested a five point response format in his seminal work, \textit{A Technique for the Measurement of Attitudes} \cite{Likert1932}.  However, Likert did not base this decision in theory and rather suggested that variations on this five-point format may be appropriate \cite{Likert1932}. Further investigation has yet to provide a consensus on the optimal number of response options comprising a Likert item \cite{Matell1971a}.  \cite{Preston2000} found that scales with four or fewer points performed the worst in terms of reliability and that seven to nine points were the most reliable. This finding is backed up by \cite{Creech2019} in their investigation of categorization error. \cite{Wu2017} demonstrated via simulation that the more points a response contains, the more closely it approximates interval data and therefore recommended an 11-point response format. 

This line of reasoning may lead one to believe that one should dramatically increase the number of response points to more accurately measure a construct.  However, just because the data may more closely approximate interval data does not mean increasing the number of response points monotonically increases the ability to measure a subject's attitude. A larger number of response options may require a higher mental effort by the participant, thus reducing the quality of the response~\cite{Bendig1953,Lee2014}. For example, \cite{Bendig1953} conducted a study that suggested that response quality decreased above eleven response options. \cite{Simms2019} also investigated the optimal number of response options and found that no further psychometric advantages were obtained once the number of response options rose above six and \cite{Lee2014} suggested based on study results that the optimal number is between four and six.

\textit{Recommendation - As a general rule-of-thumb, we recommend the number of response options be between five and nine due to the declining gains with more than ten and lack of precision with less than five. However, if the study involves a large cognitive load or lengthy surveys, the researcher may want to err on the side of fewer response items to mitigate participant fatigue~\cite{Preston2000}.}

\medskip
\noindent\textbf{Neutral Midpoint -}
Another point of contention which influences the response number of a scale is whether or not to include a neutral midpoint. Likert, with his five-point scale, included a neutral, “undecided” option for participants who did not wish to take a positive or negative stance \cite{Likert1932}.  Some argue that a neutral midpoint provides more accurate data because it is entirely possible that a participant may not have a positive or negative opinion about the construct in question. Studies have shown that including a neutral option can improve reliability in other, similar scales \cite{Courtenay1985,Madden1978,Joshi2015,Guy1997}.  Furthermore, the lack of a neutral option precludes the participant from voicing an indifferent opinion, thus forcing him or her to pick a side which he or she does not agree with. 

On the other hand, a neutral midpoint may result in users ``satisficing" (i.e., choosing the option that may not be the most accurate to avoid extra cognitive strain resulting in an over-representation at the midpoint) \cite{KrosnickJ.A.NarayanS.S.&Smith1996}. \cite{Johns2006} argue that ``{\ldots}the midpoint should be offered on obscure topics, where many respondents will have no basis for choice, but omitted on controversial topics, where social desirability is uppermost in respondents' minds."

\textit{Recommendation - We adopt the recommendation of \cite{Johns2006}, which suggests that HRI researchers utilize their best judgement based on the context of use when deciding the merits of including a neutral option in their response format.  For example, if the authors are conducting a pre-trust survey to gauge a baseline level of trust before the participant has interacted with the robot, they may want to include a neutral option since some participants, especially those unfamiliar with robots, may not truly have a good sense of their own trust in robots.  A neutral option would allow participants to present this sentiment.  However, if a survey is being utilized to assess trust after a participant has interacted with a robot, the researchers may want to remove the neutral option, arguing that participants should have developed a sense of either trust or distrust after the interaction. Nonetheless, there may be cases when ``neutral" truly is appropriate, which is why we argue in favor of researcher discretion~\cite{Johns2006}.}

\medskip
\noindent\textbf{Number of Items -}
The next point of contention we address is the ideal number of Likert items in a scale.  In his original formulation, Likert stated that multiple questions were imperative to capture the various dimensions of a multi-faceted attitude. Based on Likert's formulation, the individual scores are to be summed to achieve a composite score that provides a more reliable and complete representation of a subject's attitude \cite{Nunnally1994,Woollins1992}.

Yet, in practice it is not uncommon for a single item to be used in HRI research due to the efficiency that such a short scale provides.  Research into the appropriateness of single item scales has been extensively studied in marketing and psychometric literature~\cite{Leung2013}.  For example, \cite{Leung2013} investigated the use of a single-item scale for measuring a construct concluding that  a single-item scale is only sufficient for simple, uni-dimensional, unambiguous objects. 

Multi-item scales on the other hand are ``suitable for measuring latent characteristics with many facets.'' \cite{Rossiter2002} proposed a procedure for developing scales for evaluating marketing constructs and suggested that if the object of interest is concrete and singular, such as how much an individual likes a specific product, then a single item is sufficient.  However, if the construct is more abstract and complex, such as measuring the trust an individual has for robots, then a multi-item scale is warranted. This line of reasoning is supported by \cite{Bergkvist2007,Diamantopoulos2012,DeBoer2004}.  As to the exact number of items, \cite{Diamantopoulos2012} demonstrated via simulation that at least four items are necessary for evaluation of internal consistency of the scale.  However, as suggested by \cite{Willits}, one should be cautious of including too many items as a large scale may result in higher refusal rates.
	
\textit{Recommendation - Due to the complexity of attributes most often measured in HRI (e.g., trust, sociability, usability, etc.), we recommend that researchers in the HRI community utilize multi-item scales with at least four items.  The total number of items again is left to the discretion of the researcher and may depend on the time constraints and the workload that the participant is already facing. Because an average person takes two to three seconds to answer a Likert item and individuals are more likely to make mistakes or ``satisfy" after several minutes, we recommend surveys not be longer than 40 items~\cite{Yan2008}. Recall that this recommendation for the number of ``Likert Items" is distinct from our recommendation regarding the number of ``response options," which we recommend generally be between five and nine options, as noted previously. }

\medskip
\noindent\textbf{Scale Balance -}
The last aspect of scale design which we will discuss is that of balance.  The question of whether the items within a scale should be balanced, i.e. there should be a parity of positive and negative statements, is one less often addressed in literature. It is believed that balancing the questionnaire can help to negate acquiescence bias, which is the phenomenon in which participants have a stronger tendency to agree with a statement presented to them by a researcher. Likert \cite{Likert1932} advocated that scales should consist of both positive and negative statements. Many textbooks, such as \cite{Moule2015}, also state that scales should be balanced.  Perhaps the most compelling evidence that balance is an important factor when developing Likert scales is provided by \cite{SchumanHowardPresser1981}. The authors in \cite{SchumanHowardPresser1981} conducted a study in which they asked participants to respond to a positively worded question to which 60\% of participants agreed.  They asked the same question but rephrased in a negative way and again, 60\% of participants agreed.  This study reveals the extent to which acquiescence bias can sway participants to answer in a particular way that is not always representative of their true feelings.

One would find this evidence to be sufficiently compelling to recommend scale balance; however, this debate is not so easily settled. Recent work suggests that although including both positively and negatively worded items reduces the effects of acquiescence bias, it may have a negative impact on the construct validity (i.e., if the scale adequately measures the construct of interest) of the scale \cite{Yamaguchi1997,Quilty2009}.  This result may be due to the fact that a negatively worded item is not a true opposite of a positively worded item. Therefore, reversing the scores of the negatively worded items and summing may have an impact on the dimensionality of the scale due to the confusion that reversed items cause \cite{VanSonderen2013,Horan2009}.

\textit{Recommendation - Because of a lack of consensus and the problems arising from both approaches, we do not provide a concrete recommendation to researchers about scale balance.}

\medskip
\noindent\textbf{Validity and Reliability of Likert Prompts -}
Likert's original work states that the prompts of a Likert scale should all be related to a specific attitude (e.g., sociability) and should be designed to measure each aspect of the construct. Each item should be written in clear, concise language and should measure only one idea \cite{Nemoto2013,Likert1932}. This formulation helps to ensure the reliability (i.e., the scale gives repeatable results for the same participant) and the validity (i.e., the scale measures what is intended) of the scale.

A poorly formed scale may result in data that does not  assess the intended hypothesis. Thus, before a statistical test is applied to a Likert scale, it is best practice to test the quality of the scale. Cronbach's alpha is one method by which to measure the internal consistency of a scale (i.e., how closely related a set of items are). A Cronbach's alpha of 0.7 is typically considered an acceptable level for inter-item reliability \cite{Taber2018}. If the items contains few response options or the data is skewed, another method such as ordinal alpha should be employed \cite{Gadermann2012}.

While Cronbach's alpha is an important metric, a full item factor analysis (IFA) can be conducted to better understand the dimensionality of a scale. A scale consisting of unrelated prompts may achieve a high Cronbach's alpha for other underlying reasons or simply because Cronbach's alpha can increase as the number of items in the scale increases \cite{Goforth2016,Tavakol2011}. Furthermore, a scale can show internal consistency, but this does not mean it is uni-dimensional. On the other hand, a factor analysis is a statistical method to test whether a set of items measure the same attribute and whether or not the scale is uni-dimensional. Factor analysis thus provides a more robust metric to assess the scale quality \cite{Asun2016}.

\textit{Recommendation - Due to the complex nature of scale design, we recommend that researchers utilize well-established and verified scales provided in literature when possible. Many common constructs measured in HRI can be measured with already validated scales such as the "Trust Perception Scale" for human-robot trust or the RoSAS scale for perceived sociability  \cite{Schaefer2016metric,Carpinella2017}. This practice will reduce the prevalence of employing poorly designed scales. Otherwise, a thorough analysis of the internal consistency and dimensionality of new scales should be conducted when being employed to answer research questions. For in-depth instructions on how best to construct Likert scales from the ground up, please see \cite{Handwerker,Bala}.}

\medskip

\subsection{Statistical Tests}
Once a scale is designed and its validity statistically verified, it is important that correct statistical tests are applied to the response data obtained from the scale. Another fiercely debated topic is whether data derived from single Likert items can be analyzed with parametric tests. We want to be clear that this controversy is not over the data type produced by Likert items but whether parametric tests can be applied to ordinal data.

\medskip
\noindent\textbf{Ordinal versus Interval - }
Previous work has demonstrated that a single Likert item is an example of ordinal data and that the response numbers are generally not perceived as being equidistant by respondents \cite{Lantz2013}. Because the numbers of a scale for Likert items represent ordered categories but are not necessarily spaced at equivalent intervals, there is not a notion of distance between descriptors on a Likert response format \cite{Clason1994}. For example, the difference between "agree" and "strongly agree" is not necessarily equivalent to the difference between "disagree" and "strongly disagree." Thus, a Likert item does not produce interval data \cite{Bishop}. While it has been speculated that a large-enough response scale  can approximate interval data, Likert response scales rarely contain more than 11 response points \cite{AllenI.ElaineSeamanChristopherA.2007,Wu2017}. 

\textit{Recommendation - Because a Likert item represents ordinal data, parametric descriptive statistics, such as mean and standard deviation, are not the most appropriate metric when applied to individual Likert items.  Mode, median, range, and skewness are better to report.}

\medskip
\noindent\textbf{Parametric versus Non-Parametric - }
The question now becomes, given the ordinal nature of individual Likert items, is it appropriate to apply parametric tests to such data?  A famous study by \cite{Glass1972} showed that the F test is very robust to violation of data type assumptions and that single items can be analyzed with a parametric test if there is a sufficient number of response points. \cite{Lantz2013} demonstrates through simulation that ANOVA is appropriate when the single-item Likert data is symmetric but that Kruskall-Wallis should be used for skewed Likert item data. \cite{Creech2019} also found that skew in the data results in unacceptably high errors when the data is assumed to be interval. \cite{Meek2007} compared the use of the t-test versus the Wilcoxon signed rank test on Likert items and found that the t-test resulted in a higher Type I error rate for small sample sizes between 5 and 15. \cite{Nanna1998} made a similar comparison and also found that Wilcoxon rank-sum outperformed the t-test in terms of Type I error rates. As demonstrated by these studies, the field has yet to reach a clear consensus on whether parametric tests are appropriate, and if so when, for single Likert item data.

Likert scale data (i.e., data derived from summing Likert items) can be analyzed via parametric tests with more confidence. \cite{Glass1972} showed that the F test can be used to analyze full Likert scale data without any significant, negative impact to Type I or Type II error rates as long as the assumption of equivalence of variance holds. Furthermore, \cite{Vickers1999} showed that Likert scale data is both interval and linear.  Therefore, parametric tests, such as analysis of variance (ANOVA) or t-test, can be used in this situation as long as the appropriate assumptions hold.

\textit{Recommendation - Because studies are inconclusive as to whether parametric tests are appropriate for ordinal data, we recommend that researchers err on the conservative side and utilize non-parametric tests when analyzing Likert data.  However, we also recommend that HRI researchers avoid performing statistical analysis on single Likert items altogether.  As \cite{Carifio2007} so eloquently states, "one item a scale doth not make."  A single item is unlikely to be the best measure for the complex constructs that are of interest in HRI research as discussed in Section \ref{design}.  Therefore is best to avoid the ordinal vs. interval controversy altogether and instead perform analysis on a multi-item scale since Likert scales can be safely analyzed with parametric tests. If a researcher does choose to analyze an individual item, he or she should clearly state they are doing so and acknowledge possible implications. At the very least, it is recommended to test for skewness.}

\medskip

\noindent\textbf{Post-hoc Corrections -} The importance of performing proper post-hoc corrections and testing for assumptions are broadly applicable concerns, not specific to Likert data.  Nevertheless, they are important considerations when analyzing Likert data and are often incorrectly applied in HRI papers.

As the number of statistical tests conducted on a set of data increases, the chances of randomly finding statistical significance increases accordingly even if there is no true significance in the data. Therefore, when a statistical test is applied to multiple dependent variables that test for the same hypothesis, a post-hoc correction should be applied. Such a scenario arises frequently when a statistical analysis is applied to individual items in a Likert scale \cite{Carifio2007}. In 2006, \cite{Austin2006} conducted a study investigating whether individuals born under a certain astrological sign were more likely to be hospitalized for a certain diagnosis. The authors tested for over 200 diseases and found that Leos had a statistically higher probability of being hospitalized for gastrointestinal hemorrhage and Sagittarians had a statistically higher probability of a fractured humerus. This study demonstrated the heightened risk of Type I error that occurs when no post-hoc correction is applied.

There is controversy as to which post-hoc correction is best. \cite{Kim2015} suggests applying the Bonferonni correction when only several comparisons are performed, i.e., ten or less. The authors recommend employing a different correction such as Tukey or Scheff\'e with more than ten comparisons to avoid the increased risk of Type II errors that stems from the conservative nature of the Bonferonni correction. \cite{Nakagawa2004} suggests that researchers should, instead of performing post-hoc correction, focus on reporting effect size and confidence intervals, such as Pearson's r.

\textit{Recommendation - Because of the danger that comes with performing many statistical tests without predefined comparisons we recommend that researchers always perform the proper post-hoc corrections. Due to the increased risk of Type II error that some post-hoc tests pose, we encourage researchers to also report the effect size and confidence interval to provide a more informative and holistic view of the results.  In general, we recommend against pair-wise comparisons performed on individual Likert items for reasons already discussed.}

\medskip
\noindent\textbf{Test Assumptions -} Most statistical tests require certain assumptions to be met.  For example, an ANOVA assumes that the residuals are normally distributed (normality) and the variances of the residuals are equal (homoscedasticity) \cite{Warner2012}. Tests to ensure these conditions are met include the Shapiro-Wilk test for normality and Levene's test for homoscedasticity~\cite{CHIAROTTI2004}.  \cite{Glass1972} argues that even when assumptions of parametric tests are violated, in certain situations, the test can still be safely applied.  However, \cite{Blair1981} counters \cite{Glass1972} and contends that \cite{Glass1972} failed to take into account the power of parametric tests under various population shapes and that these results should not be trusted.

\textit{Recommendation - To navigate this controversy, we suggest that researchers err on the conservative side and always test for the assumptions of the test to reduce the risk of Type I errors. If the data violates the assumptions, and the researchers decide to utilize the test despite this, they should report the assumptions of the test that have not been met and the level to which the assumptions are violated.}



\begin{figure*}[t]
    \centering
    \includegraphics[width=0.8\textwidth]{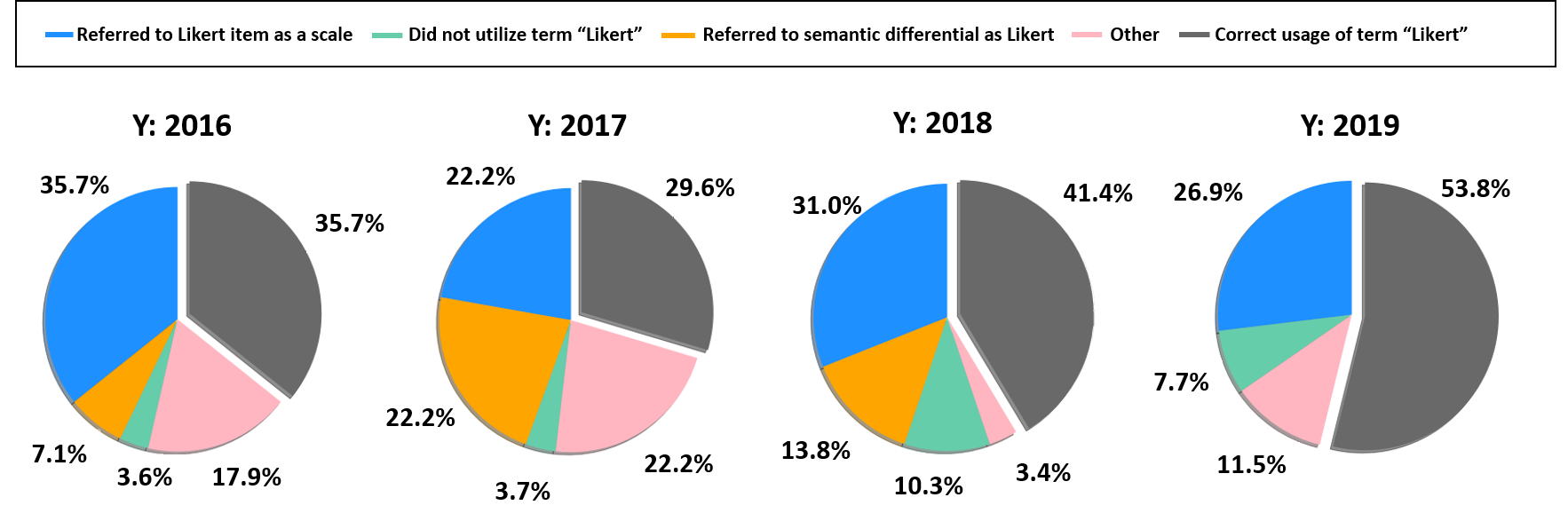}
    \caption{Common misnomer of the term "Likert Scale" within HRI Proceedings. Note: one paper in 2018 referred to a Likert item as a Likert Scale and a semantic differential scale as a Likert scale, which we counted only under the former category.}
    \label{fig:misnomer}
\end{figure*}
\section{Review of HRI Papers}
\subsection{Procedures and Limitations} We reviewed HRI full papers from years 2016 to 2019, excluding alt.HRI and Late Breaking Reports, and investigated the correct usage of Likert data over these years. We considered all papers that include the word "Likert" as well as papers that employ Likert techniques but refer to the scale by a different name. We utilized the following keywords when conducting our review: "Likert", "Likert-like," "questionnaire," "rating," "scale," and "survey." After filtering based on these keywords, we reviewed a total of 110 papers. Below we report on the following categories: 1) misnomers and misleading terminology 2) improper design of Likert scales and 3) improper application of statistical tests to Likert data.

We report on the aggregate number of papers that improperly utilized the term Likert as well as papers that improperly designed Likert scales.  Our observations also include papers that apply parametric tests to individual Likert items as well as papers that apply parametric tests to Likert scales but do not properly check for the assumptions of the test.  Furthermore, we investigate the percentage of papers that perform statistical tests to individual items that are measuring different aspects of the same attribute but do not apply appropriate post-hoc corrections. Lastly, we report the percentage of papers that calculate the mean and standard deviation associated with individual Likert items. Fig.~\ref{fig:error_overview} shows the number of papers that utilized Likert-related techniques over the years under consideration. To test if the number of papers using Likert questionnaires was correlated with the year of the proceedings, we employed a Pearson correlation coefficient test, which failed to reject the null hypothesis ($t(2) = - 0.617$, $p=0.600$) that the two factors are uncorrelated.
The test's assumption regarding normality was satisfied under the Shapiro-Wilk test, but homoscedasticity could not be tested as there is only one data point for each level (i.e., year). We reviewed each of these papers for correct practices. Our results illustrate the extent to which Likert data and scales are misused in HRI research and demonstrate the need for better practices to be employed to ensure the validity of results.  

Throughout our review, we found ourselves limited by certain papers that did not provide enough information to properly gauge whether best practices were used. We include the count of these ambiguous papers within our results under an "Other" category.  Included in this category are papers that used Likert scale questionnaires to test certain subjective metrics but did not state the number of items or other properties about the scale. This lack of detail limited our ability to determine whether their use of parametric tests were correct. In our reporting, we gave the benefit of the doubt to papers that did not report enough detail to verify the fidelity of their practices. We recommend as best practice to thoroughly report the statistical procedures used to support peer review. 

\subsection{Likert Misnomers}
First, we report on the papers that incorrectly apply the terms "Likert" or "Likert scale."  We base our analysis on the definition of Likert scale discussed in Section \ref{LikertScale}. Fig.~\ref{fig:misnomer} summarizes our findings and shows the frequency and percentages of papers that utilize each misnomer.

\medskip

\noindent\textbf{Mislabeling a Likert Item as a Likert Scale -} The phrase "Likert scale" refers specifically to a sum across a set of related Likert items, each item measuring an aspect of the same attribute. A Likert scale prompts the user to specify their level of agreement or disagreement with a set of statements (i.e., Likert items). For the term "Likert scale" to be used, the object of reference should meet these criteria. During our review, we found that references to a single Likert item as a Likert scale are ubiquitous. For example, it is common to measure an attribute of the robot by asking a participant to rate the robot according to that trait on a Likert item response scale and to refer to this single rating as a Likert scale. While such a mistake may not have an impact on the researchers' conclusion about the relevant hypothesis, it can be misleading to the reader and may imply a more robust result than what is actually achieved. Furthermore, this misnomer may imply that parametric statistical tests are appropriate when they may not be. We found that 29\% of papers labeled a Likert item as a Likert scale, and another 14\% did not provide enough information about their questionnaire for us to determine whether their application of the term was accurate.

\medskip

\noindent\textbf{Mislabeling a Semantic Differential Scale as a Likert Scale -} A "semantic continuum" consists of a set of semantic differential scales similar to how a Likert scale consists of several Likert items \cite{Verhagen2015}. A semantic continuum differs from a Likert scale in that it utilizes a bipolar scale of antonyms and measures how much of a quality a specific item has. For example, a Likert item may consist of the statement "The robot makes me sad," and the user is prompted to select how much he or she agrees or disagrees with the statement. On the other hand, a semantic differential scale will prompt the user to select how the robot makes them feel, ranging from sad to happy. Multiple semantic differential scales measuring the same attribute can be summed together to form a "semantic continuum." While a semantic continuum is appropriate to utilize in many contexts, it has important inherent differences from a Likert scale. As such, we should be careful to not mislabel one as the other. Semantic continuums are specifically useful for measuring the ``intensity and direction of the meaning of concepts" and have their own set of requirements for design as detailed in \cite{Friborg2006}. We found that an average of 7\% of papers from each year adopted this misnomer.

\subsection{Incorrect Design of Likert Scale} 
\begin{figure}[t]
    \centering
    \includegraphics[width=0.8\linewidth]{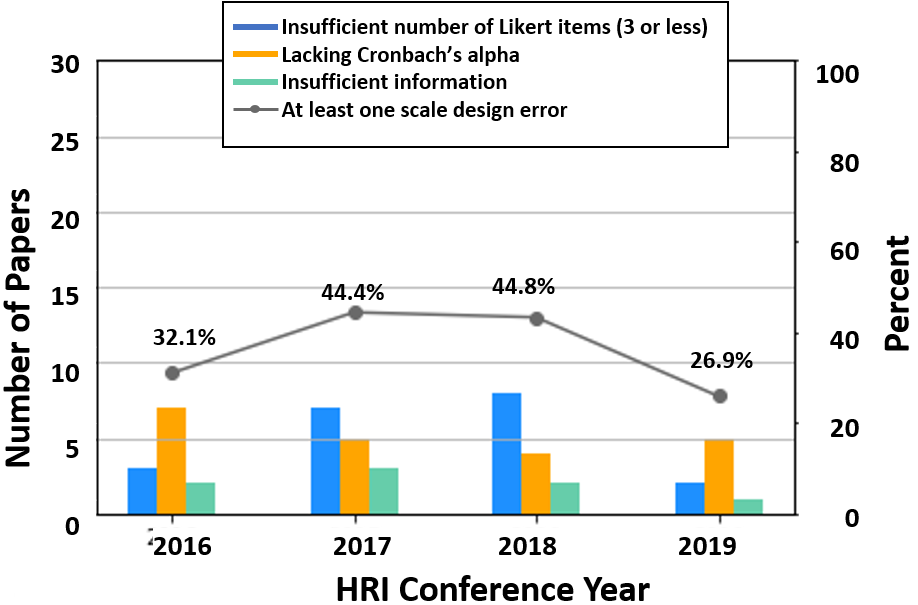}
    \caption{This figure shows the frequency of papers by year that employed improperly designed Likert scales. The percentage of papers that has at least one of these improper Likert is also reported for each year.}
    \label{fig:improper_scale}
\end{figure}

In conjunction with the improper use of the term Likert scale, we also note papers whose design or validation of a scale are questionable (see Fig.~\ref{fig:improper_scale}). Our report includes papers that utilize Likert scales with too few items, a failure to report a Cronbach's alpha, or other ambiguity within the paper's writing that could lead to disputable results. The importance of these considerations for the design of Likert scales is detailed in Section \ref{design}. We found that an average of 37\% papers had at least one of the above errors.

\subsection{Incorrect Application of Statistical Tests} 
\begin{figure}[t]
    \centering
    \includegraphics[width=0.8\linewidth]{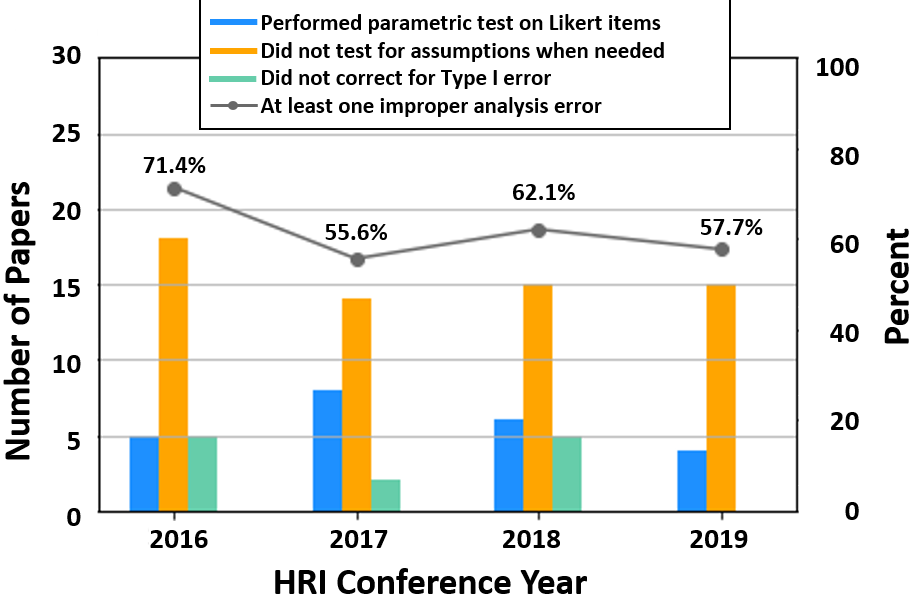}
    \caption{This figure illustrates the frequency of papers each year that incorrectly apply statistical tests on Likert data. The percentage of papers per year that incorrectly applied statistical tests is also reported.}
    \label{fig:improper_stat}
\end{figure}

In this section, we report on the recurrent ways in which statistical tests are misapplied to Likert data. We found it common for researchers to apply parametric tests to single Likert items as well as to report parametric descriptive statistics of single Likert items without stating their assumptions when doing so, both of which are not the best practice. Furthermore, papers frequently fail to check for the assumptions of parametric tests and often fail to apply appropriate post-hoc corrections. Fig.~\ref{fig:improper_stat} summarizes our findings.

\medskip

\begin{figure*}
    \centering
    \includegraphics[width=0.9\linewidth]{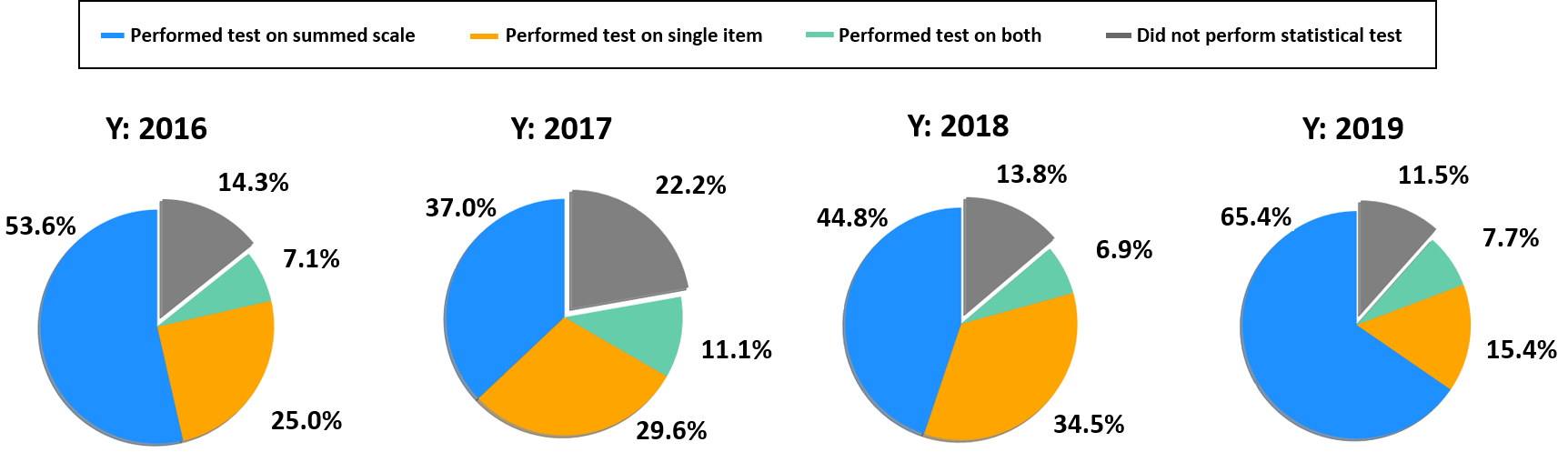}
    \caption{This figure shows the number of papers that performed statistical analysis on a Likert scale and single Likert items.}
    \label{fig:scale_vs_item}
\end{figure*}

\noindent\textbf{Application of Parametric Tests to Likert Items -} A parametric test makes certain assumptions about the distribution from which the samples were drawn.   Therefore, ANOVA, t-tests, and other parametric statistical tests are not always the most appropriate to apply to single Likert items, especially when the skew of the data is not taken into account, and their application may result in additional Type I errors. For each conference year, approximately 21\% of papers with Likert data applied parametric tests when analyzing individual Likert items without testing for skewness or detailing their assumptions when doing so. 
Fig.~\ref{fig:scale_vs_item} illustrates the number of papers that improperly analyzed single Likert items.

\medskip

\noindent\textbf{Inadequate Verification of Assumptions -} While it is not always best practice to apply parametric tests to Likert items, it is acceptable to do so with Likert scales. This allowance is because data derived from Likert scales can be assumed to be interval in nature \cite{Derrick2017}. However, most parametric tests come with a variety of assumptions that must be met before the test can be properly applied. These assumptions test whether the data in question could have been sampled, statistically speaking, from the associated underlying distribution. For example, an ANOVA assumes that the data has been drawn from a normally distributed population, and therefore, a test for normality must be performed to verify this assumption. We observed that more than 50\% of papers with Likert data from each year did not check for or report on the assumptions associated with the underlying distribution when they chose to perform a parametric test. 

\medskip
\noindent\textbf{Inadequate Post-hoc Corrections -}
In general, post-hoc corrections may be performed when several dependent variables are testing the same hypotheses or when multiple statistical tests are performed on the same variables. For example, if a researcher con-ducts a statistical test on each individual item in a Likert scale, a correction should be applied since this is an example of testing several dependent variables that are assessing the same hypothesis. Furthermore, the chance of a Type I error increases as the number of dependent variables being tested increases. On average, we found that 11\% of papers with Likert data did not account for this increased likelihood of family-wise error when they chose to perform a statistical test on individual items related to one hypothesis. For the papers that reported p-values, we performed a Bonferroni correction in order to determine the validity of the paper's result. On average 40\% of the results reported in each of these papers were not significant after the adjustment. This lack of significance \ul{does not} mean that the papers' conclusions are incorrect, considering the conservative nature of the Bonferonni correction. Rather, this lack suggests findings should be re-examined with proper methods.

\medskip

\noindent\textbf{Incorrect Reporting of Descriptive Statistics -} Another common practice we found is reporting the mean and standard deviation of individual Likert items. An average of 31\% of papers with Likert data from each year reported their Likert item results in this descriptive manner, most commonly through visual bar graphs. This practice is unhelpful as Likert items are ordinal data without a concept of mean or standard deviation in ordinal data. Appropriate descriptive metrics are median, mode, and range.

\section{Discussion}

Our review of four years of HRI proceedings shows that nearly all relevant papers committed at least one error that could raise questions about the inferences drawn from the data. The overall trend observed between the four years does not appear to improve, leading us to believe that a call to action is warranted.

Specifically, we should seek to avoid misapplying the term Likert scale, design scales with an appropriate number of items, and test for the assumptions of the statistical analyses being applied. An in-depth review of HRI proceedings shows that the use of the term Likert scale has taken a looser connotation, as we found that roughly half of all the misnomer errors were from papers describing the response scale as a Likert scale. With respect to certain papers designing their own Likert scale for a specified metric, 18\% of papers have less than four items to measure a complex construct. 

Our review also shows that a large number of papers do not properly perform statistical analysis on Likert scales. Because a Likert scale is a summation across Likert items, the resulting values approximate interval data, which allows for parametric tests to be performed. However, for parametric tests to be applied, the assumption of the underlying distribution must still be tested for; and yet, 56\% of papers we reviewed did not confirm this key assumption. 


Finally, our analysis does not refute the conclusions of any HRI paper. Our key take-away is that we should strive for better practices so that we can be more confident in the conclusions we draw from the data. Our findings also bolster the recent support of reproducibility studies as full contributions in the field of HRI.

\section{Theses}
We list our recommendations to the HRI community based upon our review of the psychometric literature and in light of our findings of current HRI practices. Bold typeface is used for points made in response to the most common Likert scale issues.
\begin{itemize}
    \item \textbf{Referring to a response scale as a Likert scale is a misnomer.} Instead, use ``response format" or "response scale" when discussing the value range and reserve the term Likert scale for when referring to the entire set of items.\vspace{2pt}
    \item Questions within a Likert scale should measure the various aspects of one and only one subjective attitude or construct.\vspace{2pt}
    \item Likert scales should be checked for internal consistency and uni-dimensionality to ensure their reliability and validity.\vspace{2pt} 
    \item \textbf{A single Likert item should not be a sole metric for measuring a multi-faceted construct, as one statement is not generally sufficient to fully capture a complex attitude.} We recommend having at least four items.\vspace{2pt}
    \item We encourage utilization of well-developed and validated Likert scales, e.g.~RoSAS and SUS, when possible \cite{Carpinella2017,Brooke1996}.\vspace{2pt}
    \item \textbf{The ordinal nature of Likert item data should be considered when selecting an appropriate statistical test.}\vspace{2pt} 
    \item It is important to systematically check for and satisfy all assumptions of the statistical tests being applied to the data. \vspace{2pt}
    \item Experiments should be replicable: thorough detail should be provided regarding design and testing of Likert items, scales.\vspace{2pt}
    \item \textbf{If there is more than one dependent measures supporting a single hypothesis, a correction to account for Type I error should be applied.}
\end{itemize}

\section{Conclusion}
 A majority of published HRI papers rely on Likert data to gain insight into how humans perceive and interact with robots, leading Likert questionnaires to be a fundamental part of HRI studies. In this paper, we reviewed HRI proceedings from 2016-2019 and reported aggregate results of the improper use of Likert scales. Furthermore, we explored the implications of these infractions via a literature review on simulations and studies focused on incorrect design and statistical testing of Likert scales and associated data. While it is encouraging that the observed trends of the papers containing problematic usage of Likert scales and data has not increased over the last four years, it is our belief that we as a community should strive for better practices. The authors of this paper are included in this call to action. It is our hope that our recommendations are taken into consideration and that HRI researchers, authors, and reviewers employ best practices when addressing Likert data.   
 
\begin{acks}
We thank Ankit Shah for his statistical insights and support. This work was supported by institute funding at the Georgia Institute of Technology and NSF ARMS Fellowship under Grant \#1545287.
\end{acks}

\bibliographystyle{ACM-Reference-Format}
\bibliography{HRIPapers.bib}


\end{document}